\documentclass[epjCONF, onecolumn]{svjour}
\usepackage{graphics}
\usepackage[varg]{txfonts} 
\usepackage[latin1]{inputenc}

\session-title{Conference Title, to be filled}
\begin{document}
\title{The Milky and its Gas: Cold Fountains and Accretion}
\author{Felix J. Lockman\inst{1}\thanks{The National Radio
    Astronomy Observatory is operated by Associated Universities,
    Inc., under a cooperative agreement with the National Science
    Foundation. \email{jlockman@nrao.edu}} }
\institute{National Radio Astronomy Observatory, P.O. Box 2, Green   Bank, WV 24944 USA}
\abstract{
The Milky Way is acquiring gas from infalling 
high-velocity clouds.  The material enters a disk-halo interface
that in many places is populated with HI clouds that have been ejected
from the disk through processes linked to star formation.  The Smith
Cloud is an extraordinary example of a high-velocity cloud that is  bringing $>10^6$ M$_{\odot}$ of
relatively low metallicity gas into the Milky Way.  It may be part of
a larger stream, components of which are now passing through the disk.
} 
\maketitle
\section{Introduction}
\label{intro}
The disk-halo transition zone in spiral galaxies is a lively place.  
Energetic events associated with star formation can lift
gas several kpc into the halo \cite{HowkSavage,Pidopryhora}.
Some spirals have an associated population of 
extra-planar HI clouds  that are probably being accreted from
elsewhere  \cite{Thilker,Grossi}.  
  Gas stripped from dwarf  galaxies can add to the environment
 as well.  These phenomena give many spirals a
ragged outer HI envelope with a variety of kinematics beyond
simple circular rotation \cite{Sancisi}.   Recent studies
of the Milky Way have provided new insight into these processes
that shape the evolution of galaxies.

\section{HI Clouds in the Disk-Halo Interface}
\label{sec:1}

The discovery of a population of discrete neutral hydrogen clouds
extending outward from the disk several kpc into the halo raises
a number of questions, among them whether the clouds are recent condensations from a
hot galactic halo, or a component of the disk ISM ejected from the
plane \cite{FJLclouds,Stil,Stanimirovic2006}.  The ``disk-halo''
clouds are rotating with the Galactic disk, and their numbers decrease
exponentially  with distance from the 
plane, but their cloud-cloud velocity dispersion, if
representative of an isotropic random motion, is nearly an order of magnitude
too small to account for the clouds' scale height.

Recently,  several hundred of these clouds were analyzed 
using data from a uniform survey of two regions
symmetric about the line of sight to the Galactic Center -- one in the
first quadrant of longitude (QI) and another in the fourth (QIV) \cite{Ford2008,Ford2010}.  The
observations were compared with simulations that properly account for
various biases including distance errors and confusion.  The derived vertical
distribution of the disk-halo HI clouds  shown in Figure 1 
reveals a major anomaly: 
  while the cloud-cloud velocity dispersion (and indeed all
properties of individual clouds) is identical in
the two quadrants, both the number of clouds and their scale height
differs significantly.  After ruling out possible systematic effects,
Ford et al. 
\cite{Ford2010} conclude that the cloud numbers and scale-height are
related to star formation: 
   the location chosen for study in Quadrant I coincidentally includes 
   an area with many HII regions, while the
 longitude-symmetric area in Quadrant IV lies between major spiral
 arms.  

 The disk-halo clouds may thus be  created through 
 the breakup of large superbubbles, indeed, some clouds are arranged in
 long chains  that stretch out of the disk, perhaps
 analogs to the dust pillars of NGC 891 \cite{HowkSavageNGC891}. How
 these might evolve into discrete HI clouds is not known.  
  In the classic picture of the galactic fountain,
hot gas heated by supernovae rises buoyantly then cools, condenses and
 returns to the disk \cite{HouckBregman}.  The disk-halo clouds appear more likely to have
 been lofted away from the plane in structures that exist for perhaps
 10 Myr before breaking up \cite{Pidopryhora}.  
They may be the products of a galactic fountain, but the flow is in
cool  neutral rather than hot ionized gas.

\begin{figure}
\resizebox{0.60\columnwidth}{!}{%
\includegraphics{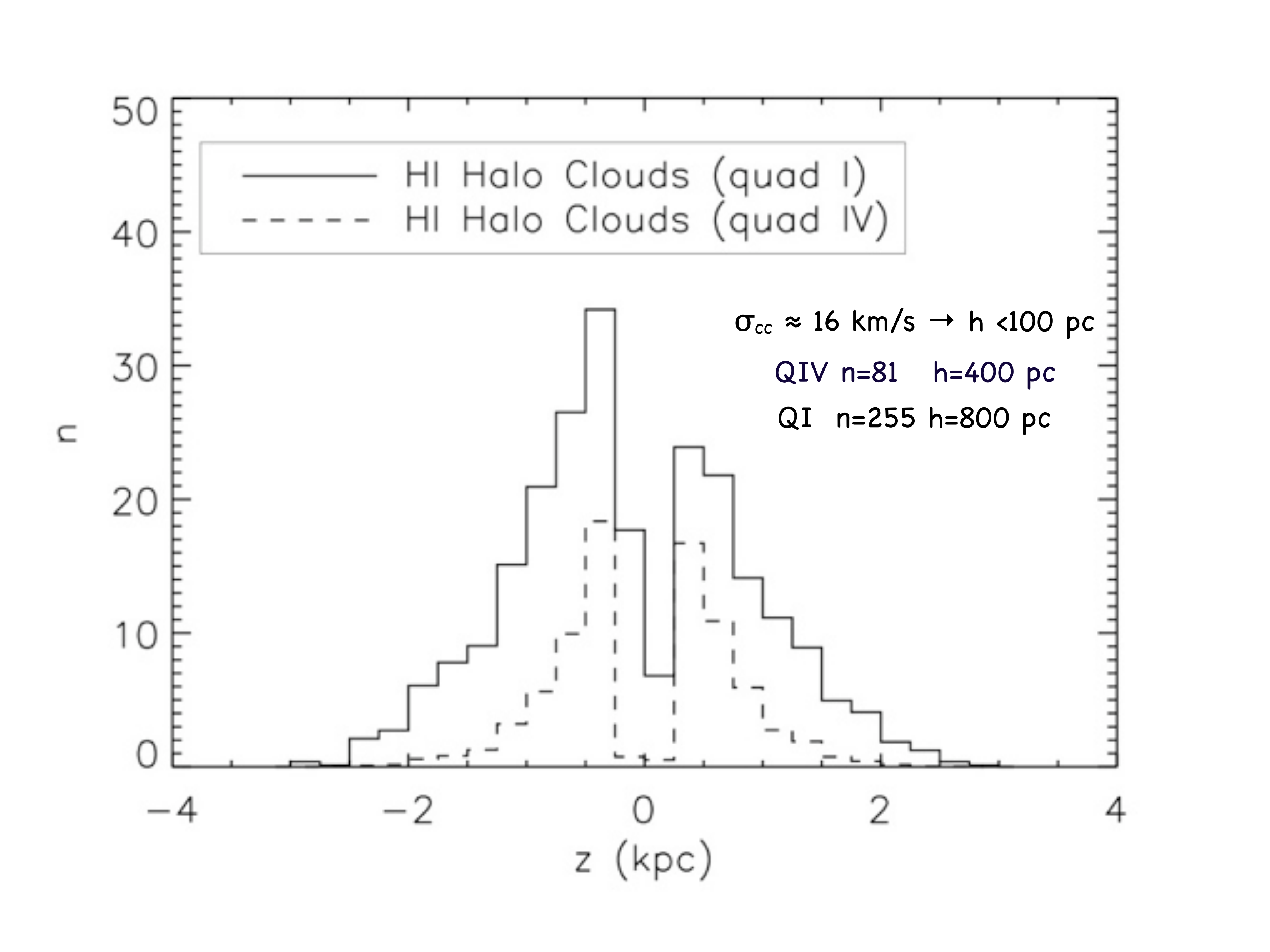} }
\caption{Observed number of disk-halo HI clouds as a function of
  distance from the Galactic plane for samples in identical volumes of
  the First and Fourth Galactic quadrants (adapted
  from Ford et al. \cite{Ford2010}).  QI has more clouds and a larger scale height than QIV, 
while the cloud-cloud velocity
  dispersion is identical and  too small to account for even the smaller scale
  height.  
The decrease in cloud numbers near the plane
  results from confusion.
}
\label{fig:1}       
\end{figure}

\section{Gas Flows into the Milky Way}
\label{sec:2}

It has long been suggested that the high-velocity HI clouds (HVCs)
that cover much of the sky are accreting from a hot halo that is
either a relic of the formation of the Milky Way or is powered by a
galactic fountain.  Some HVCs might also be the ISM stripped from
satellite galaxies, although the Magellanic Stream is the only
convincing local example.  The low metallicity of many HVCs precludes
their formation in a galactic fountain, but the origin and fate of
most HVCs is still unknown \cite{WakkervanWoerden,Wakker,Shull}.
Recent observations of the HVC known as the ``Smith Cloud''
\cite{Smith} show that it is an exceptional object, one that might
give general insight into the way that the Milky Way continues to
acquire gas.

\begin{figure}
\resizebox{0.75\columnwidth}{!}{%
\includegraphics{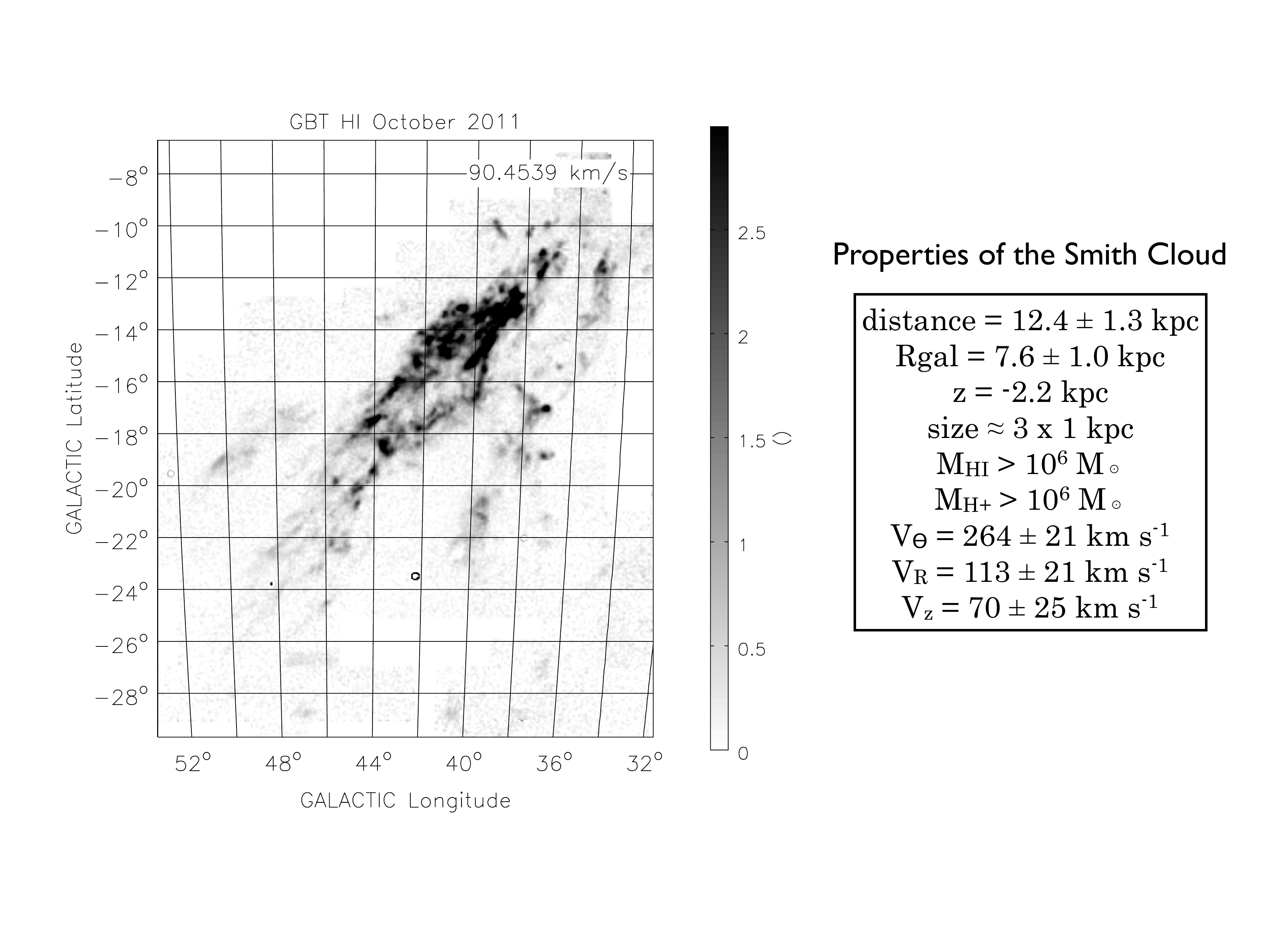} }
\caption{HI map of the Smith Cloud at V$_{\rm LSR} = 90$  km s$^{-1}$
  from new GBT data.  It is moving to
the upper right, interacting with the Milky Way halo, and 
being stripped of its outer parts.  
}
\label{fig:2}       
\end{figure}

\subsection{The Smith Cloud}
\label{sec:2.1}

Figure 2 shows an HI channel map of the Smith Cloud  from a new survey made with the
100 meter Green Bank Telescope (GBT) of the NRAO at an angular
resolution of 9', 
 observed and reduced as described by \cite{Boothroyd}.  
Properties of
the Cloud are summarized next to the Figure \cite{FJLSmith,Hill}.   This HVC 
is unique in that both its distance and its full space motion are
known, and while it has only recently begun to be studied in detail,
what we have learned already is very informative.

The Smith Cloud has a total velocity $\sim 300$ km s$^{-1}$, so  it is
gravitationally bound to the Milky Way.   The 
largest velocity component is in the direction of Galactic rotation.  It lies interior to the Solar Circle at
R$_{gal} =7.4$ kpc, and is moving upwards toward the Galactic plane,
which it will cross in about 30 Myr at R$_{gal} \approx 11$ kpc
from the Galactic center.  The tip of the cloud at $z = -2.2$ kpc is
closer to the plane than the top of some superbubbles
\cite{Pidopryhora}.  Its morphology and detailed kinematics show that
it is currently being disrupted and parts of it decellerated to the velocity of
the halo gas it is encountering \cite{FJLSmith}.  Optical emission
line measurements indicate that it contains at least as much ionized
gas as neutral gas,
and it  has a metallicity below solar \cite{Hill}.  It has
exactly the properties needed to fuel future star formation in the
Milky Way and account for the chemical evolution of the disk
\cite{Pagel}.  It is not known to have any stars.

 Nichols \& Bland-Hawthorn \cite{Nichols} have modeled the cloud
as the baryonic component of a $\sim10^8$ M$_{\odot}$ dark matter
subhalo, which had lost at least half of its dark and
baryonic mass  to the Milky Way in an earlier passage through the
disk.  They estimate that the progenitor had an initial mass that
was an order-of-magnitude higher than we observe today.

\begin{figure}
\resizebox{0.65\columnwidth}{!}{%
\includegraphics{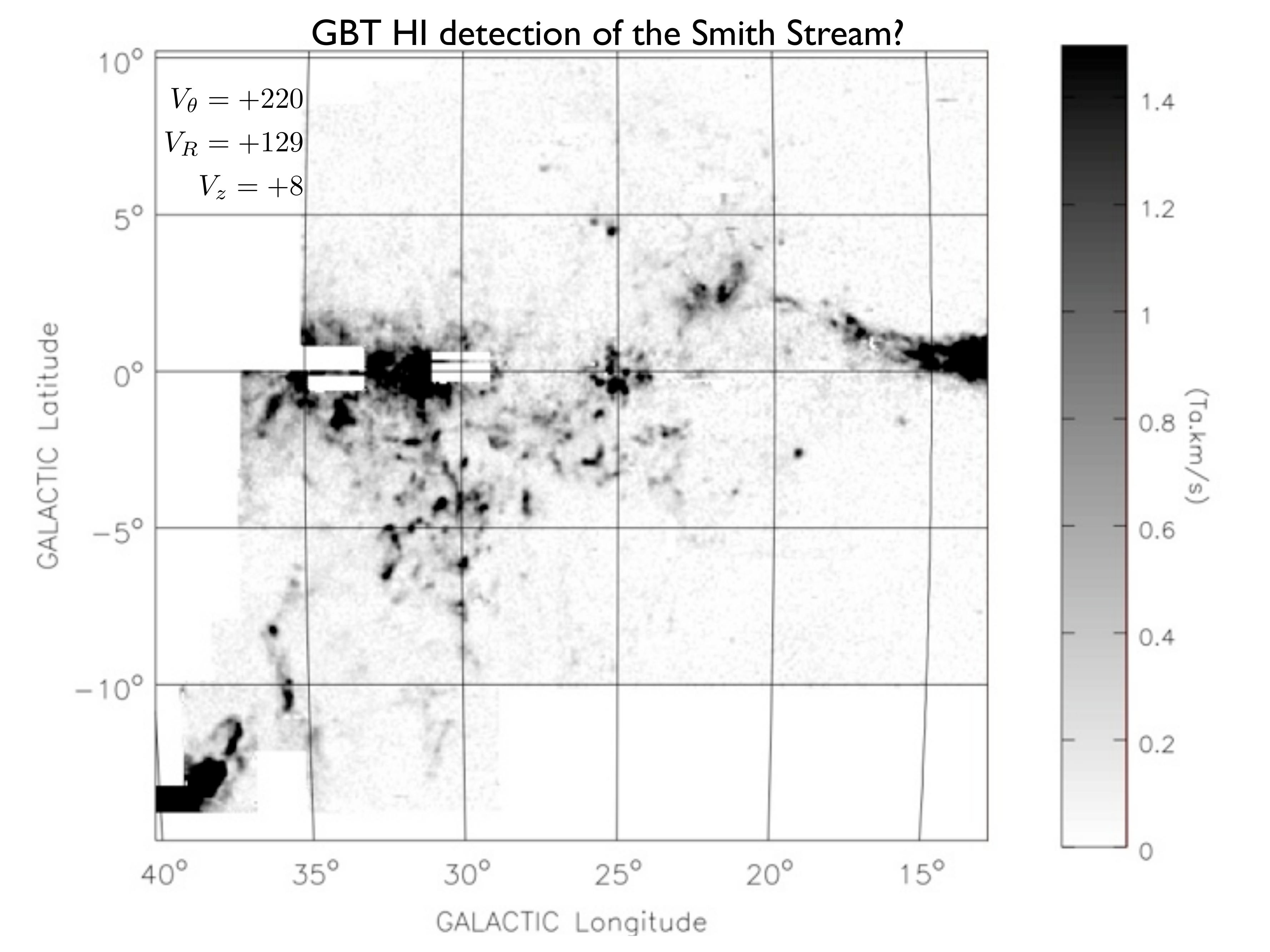} }
\caption{New GBT HI observations of the area  ahead of the path
 of the Smith Cloud (visible in the lower left).   This ``channel
  map'' shows HI at the velocity calculated from the projection
  of the velocity components given in the upper
  left.  There is a chain of HI clouds exactly along the predicted
  path and with kinematics simliar to that 
  of the Smith Cloud, possibly components of the Cloud that are
 already passing through the Milky Way  disk.  }
\label{fig:3}       
\end{figure}

\subsection{The Smith Stream?}

Is the Smith Cloud a solitary object or might it be part of a
larger structure?  To examine this we have recently observed a large
region of the sky ``ahead'' of the Smith Cloud in the 21cm line with
the GBT.  Figure
3 shows the HI map that results if at each pixel we display the HI
intensity  at the velocity which is the projection 
of $V_{\theta}, V_{R}, V_{z} = +220, +129, +9$ km s$^{-1}$,
velocity components not too different from that derived for 
the Smith Cloud.  The tip of the Smith Cloud shows up clearly in the
lower left of the Figure, and in most pixels there is no HI emission at the
projected velocity.  However,  the projection reveals a chain of clouds
stretching along the trajectory of the Smith Cloud through the
Galactic plane.  
It is possible that these are segments of an extended stream of gas, all
following a similar trajectory, of which the Smith Cloud is only the
brightest part, and as yet the least disrupted.  If so, the ``Smith
Stream'' would have an extent of more than 6 kpc.

\section{Future Work}

The model by Nichols \& Bland-Hawthorn offers one possible explanation
for the properties of the Smith Cloud; other work has suggested that
HVCs without dark matter halos are disrupted significantly before they
can reach the disk \cite{Heitsch}.  This is clearly an area that needs
more investigation, and more detailed observations of the Smith
Cloud may be able to provide information on the disruption process.
The metallicity of the Cloud also needs direct measurement, as current
estimates have a large uncertainty \cite{Hill}.  Work is likewise
needed on the evolution of superbubbles and the origin of the
disk-halo clouds.  The Smith Cloud is now traversing the  area
populated by the disk-halo clouds and it would not  be surprising to
see signs of its interaction with a lumpy, cloudy, Galactic halo.

\end{document}